\documentstyle[prd,aps]{revtex}

\def\be {\begin{equation}}
\def\ee {\end{equation}  }
\def\beq{\begin{eqnarray}}
\def\eeq{\end{eqnarray}  }
\def\bi {\begin{itemize} }

\def\ei {\end{itemize}   }
\def\RE {I\kern-6pt R    }
\def\Z  {Z\kern-13pt Z   }
\def\be {\begin{equation}}
\def\ee {\end{equation}  }
\def\beq{\begin{eqnarray}}
\def\eeq{\end{eqnarray}  }

\def\eeq{\end{eqnarray}  }

\input epsf

\begin{document}
\draft

\twocolumn[\hsize\textwidth\columnwidth\hsize\csname
@twocolumnfalse\endcsname

\title{Singularity Formation in  2+1 Wave Maps}
\author{James Isenberg}
\address{Department of Mathematics\\
         University of Oregon,
         Eugene, OR 97403}
\author{Steven L. Liebling}
\address{Theoretical and Computational Studies Group\\
     Southampton College, Long Island University,
     Southampton, NY 11968}

\maketitle

\begin{abstract}
We present numerical evidence that singularities form in finite time
during the evolution of 2+1 wave maps from spherically equivariant initial
data of sufficient energy.
\end{abstract}


\vskip2pc]

\section{Introduction}

While it has been shown that wave maps on a $1+1$ dimensional Minkowski
spacetime base evolved from smooth initial data exist for all time
~\cite{1dim1,1dim2}, and that those on an $m+1$ $(m\geq 3)$ Minkowski
spacetime base can blow up in finite time ~\cite{singular}, global
existence for the $2+1$ case remains as yet unresolved. Scaling
considerations identify $2+1$ as the critical dimension for wave maps,
and so there is considerable interest in determining if indeed $2+1$ wave
maps developed from smooth initial data can become singular in finite
time or not. Here, we describe numerical work which strongly supports the
contention that, at least for some sets of smooth initial data, they can.

There are special classes of $2+1$ wave maps for which global existence
has been shown to hold: a) spherically equivariant wave maps with
convex~\cite{convex}, or slightly more general targets~\cite{general},
b) spherically symmetric wave maps with compact targets (plus a further
technical
condition on the target)~\cite{compact}, c) general wave maps (general target)
with sufficiently small energy.

Not included in any of these three classes are spherically equivariant
wave maps from $2+1$ Minkowski spacetime into the round two-sphere with
initial data of arbitrary energy. Shatah and Struwe~\cite{struwe} have
conjectured
that singular behavior should be found in this class. Our numerical
results reported here strongly support the validity of this conjecture.

We examine one-parameter families of data, with small values of the
parameter corresponding to small energy data and therefore global
existence, and with large values of the parameter corresponding to data
possibly leading to singularity formation. One might hope to find
especially interesting wave map development for data at or near the
transition between small and large values. While this sort of ``critical"
behavior has been seen and studied in $3+1$ wave maps~\cite{liebling,bizon3+1},
we have not found nearly as clear an indication of universal critical behavior
for the present $2+1$ case. This criticality issue needs further study,
and is not treated in this paper.  Here, our focus is on numerical
evidence for singular wave map evolution from regular initial data.

We note that our studies of singularity formation in $2+1$ wave maps have
been carried out independently of the work of Bizo\'n, Chmaj, and
Tabor~\cite{bizon2+1}
using numerical algorithms which differ from theirs. However their
results and ours agree substantially.

\section {The Equations}

Generally a wave map is defined to be a map $\phi^A$ from a spacetime
(the ``base") into a Riemannian geometry (the ``target"), with $\phi^A$
a critical point for the action
\be
S[\phi] = \int_{M^{m+1}} \eta^{\mu \nu} g_{AB}\left(\phi\right)
                    \left( \partial_\mu \phi^A \partial_{\nu} \phi^B \right)
\ee
where $g_{AB}$ is the Riemannian metric on the target manifold $N^n$, and
$\eta^{\mu\nu}$ is the (inverse) Lorentz-signature metric on the
spacetime $M^{m+1}$. The Euler-Lagrange equations for this action take the
form
\be
    \partial^\mu \partial_\mu \phi^A
  + \Gamma^A_{BC} \partial_\mu \phi^B \partial^\mu \phi^C = 0
\label{eq:PDE}
\ee
where $\Gamma^A_{BC}$ represents the Christoffel symbols corresponding to
the target metric $g_{AB}$. This is a semilinear hyperbolic PDE system for
$\phi^A$. We note that for certain targets, wave maps are known to
physicists as ``nonlinear sigma models."

As noted above, the case of primary interest here is $2+1$ Minkowski
spacetime for the base and the round two sphere for the target. In this
case, the wave map PDE system (\ref{eq:PDE}) may be rewritten in the
following form
\be
\Box \phi^a + \left( \partial_\mu\phi^b  \partial^\mu\phi^c
\right) \delta_{bc}
\phi^a = 0.
\label{eq:wave}
\ee
where the indices $a, b,c$ take the values $\{1,2,3\}$ (indexing the
ambient Euclidean 3-space for the target two sphere), and $\delta_{bc}$
is the metric for this ambient space. If we now impose the condition that
the maps $\phi^a$ be spherically equivariant with angular wrapping number
k, and write $\phi^a(r,\theta,t)$ in the ``hedgehog" form
\be
\phi^a = \left( \begin{array}{c}
                \sin \chi(r,t) \sin k \theta \\
                \sin \chi(r,t) \cos k \theta \\
                \cos \chi(r,t)
                \end{array} \right)
\label{eq:hedgehog}
\ee
where $r$ is the radial distance from the origin and $\theta$ is the
azimuthal angle, then the wave map PDE system~(\ref{eq:PDE}) reduces to the
single
equation
\be
\ddot \chi = \frac{1}{r} \left( r \chi' \right)' - \frac{k^2\sin 2 \chi}{2r^2}
\label{eq:eom}
\ee
where a prime and a dot denote partial derivatives with respect to $r$ and $t$
respectively. Thus the study of the Cauchy problem for $2+1$ spherically
equivariant ($k$-wrapped) wave maps into the round two sphere focuses on
finding solutions $\chi(r,t)$ to Eq.~(\ref{eq:eom}) with regular initial data
$\chi(r,0)$, $\dot\chi(r,0)$. Note that regularity at $r=0$ requires that
we set $\chi(0,t)=0$ for all $t$.

While it may be interesting to examine if there is any variation
of the behavior of solutions for  wrapping numbers $k$ greater than one,
we restrict our attention here to the single angular wrapping case $k=1$.

As for any field theory on Minkowski space, there is a divergence-free
stress-energy tensor $T_{\mu\nu}$ associated with wave maps. From
$T_{\mu\nu}$, we obtain the energy density function for spherically
equivariant wave maps

\be
\rho(r,t) = \frac{1}{2} \left[ \dot \chi^2 + \left(\chi'\right)^2 \right]
             + \frac{\sin^2 \chi}{2r^2}
\label{eq:rho}
\ee
whose integral

\be
E(t) = \int_r \rho(r,t)~r~dr
\label{eq:energy}
\ee
is conserved (ie, $E(t)=E(0)$ for all $t$). The energy is a useful monitor
of numerical accuracy, as discussed below.

\section {Numerical Studies of Singularity Formation}

Our numerical experiments consist of specifying parametrized families of
initial data $\{\chi_\lambda(r,0), \dot \chi_\lambda (r,0)\}$ and
numerically evolving a number of sets of such data in each family. A
typical family--one of the simplest--is the approximately ingoing
Gaussian pulse
\beq
     \chi(r,0) & = & A e^{-\left(r-R_0\right)^2/\delta^2} \cr
\dot \chi(r,0) & = & \chi'(r,0).
\label{eq:id}
\eeq
This family has three parameters $A$, $R_0$, and $\delta$, with the most
important one for our discussion being the scale parameter $A$. Note that
the ingoing character of these solutions, which results from the choice of
$\dot\chi(r,0)$,  minimizes outer boundary effects. Note also that while,
analytically, $\chi(0,0)$ is not zero,
for the choices of $R_0$ and
$\delta$ which we make, we can force $\chi(0,0)$ to be zero and retain 
smoothness to within numerical accuracy.

We evolve using a second order finite difference approximation to
Eq.~(\ref{eq:eom}).  We use an iterative Crank-Nicholson scheme
implemented with RNPL~\cite{rnpl}, and  also make use of the adaptive mesh
framework developed by Choptuik~\cite{choptuik}. We have verified that
the code generates solutions which converge quadratically in the grid
spacing and conserve energy. In arguing that we are indeed generating
singularities, we will discuss the convergence and energy conservation
tests in more detail below.

\begin{figure}
\epsfxsize=8cm
\centerline{\epsffile{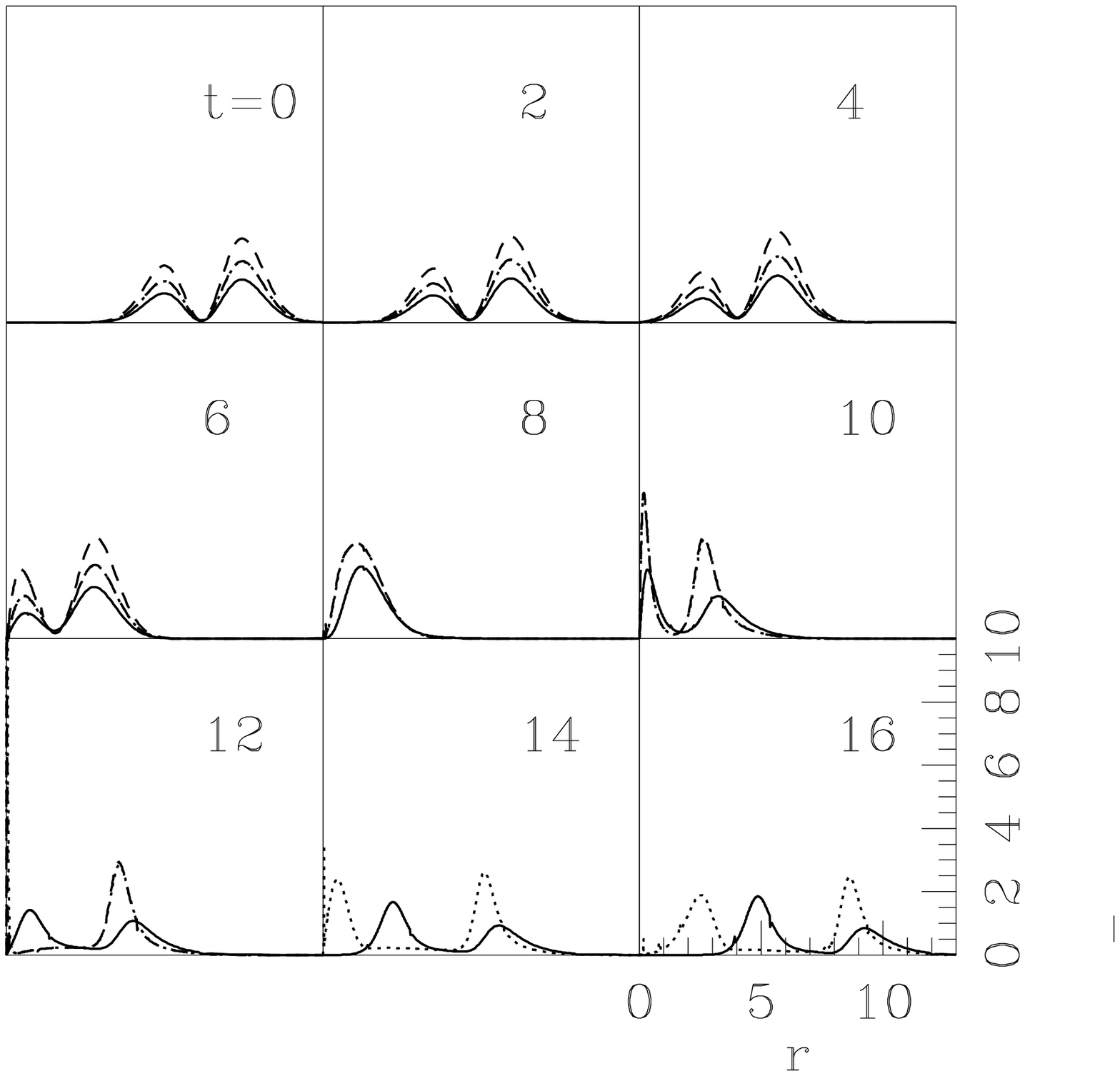}}
\caption{Snapshots of the energy densities (times $r$) for a single family
         of initial data
         with varying amplitude. At $t=0$ the initial energy densities
corresponding
         to ingoing Gaussian initial data ($R_0=8$, $\delta=2.3$) are shown.
         Supercritical ($A=1.4$) is shown long-dashed (red) which exists only
         until $t=6$. Sub-critical ($A=1.0$)
         is shown solid (magenta). Two near-critical evolutions are also shown:
         slightly subcritical ($A=1.19$) shown dotted (blue) and slightly
supercritical
         ($A=1.195$) shown short-dashed (green). The two near-critical
evolutions
         coincide at the scale of this graph until $t=12$ after which we
cannot compute
         the apparently singular supercritical solution. The energy densities
         reached by the supercritical solutions extend significantly off
the scale
         of this graph.  }
\label{fig:evolve}
\end{figure}

For a general set of ingoing Gaussian pulse data, regardless of amplitude,
the wave map evolution has the pulse maximum and energy density maximum
initially
moving inward (decreasing $r$). For small (subcritical) values of $A$,
this inward motion of the maximum proceeds for a finite time, after which
the maximum ``bounces"  away from the origin and begins to move outward
(see Fig.~\ref{fig:evolve}). There is a general dispersal of the energy
density; and
for large $t$, there is very little energy density remaining near the
origin.

For large (supercritical) values of $A$, the behavior of the evolving
wave map is qualitatively the same initially. However, rather than
bouncing away from the origin, the maxima for supercritical data
continue to approach the origin (Fig.~\ref{fig:evolve}), with the
concentration of
energy around the origin appearing to grow without bound.  As the energy
density and the gradient of the function $\chi$ grow very large at the
origin, the numerical evolution inevitably becomes unable to resolve the
gradient, and the solution becomes sufficiently non-smooth to cause the
numerical evolution  to stop. If this accumulation is indeed a
singularity forming, there is no hope for the numerical evolution to
resolve it, being itself of finite resolution. The task then is to
examine the behavior of the numerical solution up to this point.

Before doing so, we first discuss a couple of standard tests of a
numerical solution. We let $\chi(r,t)$ be some solution to the
(continuum) partial differential equation~(\ref{eq:eom}) and let $\tilde
\chi_h(r,t)$ be the solution to a discrete form of that equation, for
corresponding initial data, on a grid spacing $h\equiv\Delta r$. The hope
is  that, as the grid spacing $\Delta r$ gets smaller, the
solutions to the discrete equation generated by the evolution code
converge to the  solutions of the PDE,
$\tilde
\chi(r,t)
\rightarrow
\chi(r,t)$. Because in general the explicit solutions to the PDE are
unknown, we instead consider a series of numerical solutions on grids of
increasing resolution, say $\tilde \chi_{4h}$, $\tilde \chi_{2h}$,
$\tilde \chi_h$. If these are to converge to the PDE solution, then
they must converge themselves. To examine this convergence, we define a
convergence factor ($Q$) as follows
\be
$Q$ \equiv \frac{ |\tilde \chi_{4h} - \tilde \chi_{2h}|_2 }
                      { |\tilde \chi_{2h} - \tilde \chi_{ h}|_2 },
\label{eq:cvf}
\ee
where the norms are the $l_2$ norm.
For these solutions to converge, the difference between solutions
for increasing resolution must decrease and hence $Q$ must
be greater than one. For second-order schemes, $Q$ is
expected to be~$4$.

Another common test of numerical accuracy focuses on the degree to which
energy is conserved by the numerical evolution. The evolution governed by
the PDE~(\ref{eq:eom}) does conserve energy; the question is whether this
remains
true for the numerical evolution. Letting $E_{num}(t)$ denote the
energy  calculated from the numerical solution at time $t$ (on the finite
grid), and setting $\Delta (t) \equiv \ln \left| \frac{\
E_{num}(t)-E_{num}(0)}{E_{num}(0)}\right|$, we monitor $\Delta(t)$ for
different choices of grid spacing. The expectation is that $\Delta(t)$
should decrease with increasing resolution; if we observe this, our
confidence in the accuracy of our numerical solution is enhanced.

\begin{figure}
\epsfxsize=8cm
\centerline{\epsffile{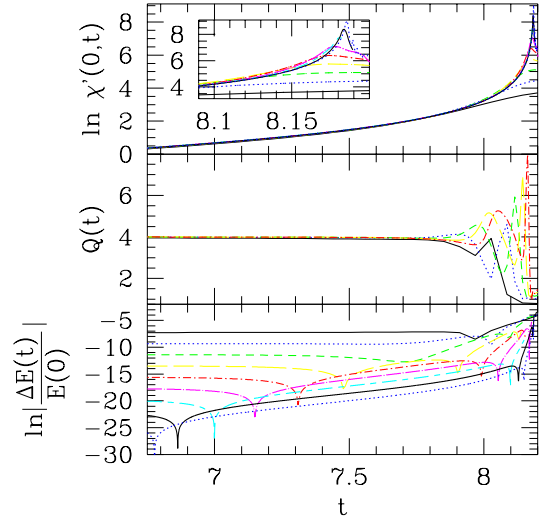}}
\caption{Results of a super-critical evolution for an initially ingoing
         Gaussian pulse ($A=2$, $R_0=10$, $\delta=2.3$,
         $R_{\rm max}=30$). The results are shown for increasing
         resolutions $n=2^8$ (solid, black), $n=2^9$ (dot, blue),
         $n=2^{10}$ (short dash, green), $n=2^{11}$ (long dash, yellow),
         $n=2^{12}$ (dot-short dash, red), $n=2^{13}$ (dot-long dash, magenta),
         $n=2^{14}$ (short dash-long dash, cyan),
         $n=2^{15}$ (solid, black),
         and $n=2^{16}$ (dot, blue),
         where $h = R_{\rm max} / n$.
         The top frame shows the rapid
         growth of $\chi'(0,t)$ near the time of the blow-up ($t \approx 8$).
         The middle frame shows the convergence factor (defined
         in Eq.~(\ref{eq:cvf})). Factors greater than one indicate
         convergence. The bottom frame shows the change in energy
         with respect to the initial energy. As the resolution
         increases, so does the level of energy conservation.
         }
\label{fig:super_test}
\end{figure}

In Fig.~\ref{fig:super_test}, we show the evolution in time of three
quantities--$\ln \chi'(0,t)$, $Q(t)$, and $\Delta(t)$--for numerical
runs of supercritical ingoing Gaussian pulse data, done with nine
different grid spacings.  In the top frame, we show the
behavior of the derivative of $\chi$ at the origin as a function of time.
The figure shows that as the pulse travels inward, the derivative
increases. Until just before $t=8$, all the resolutions show the same
behavior as would be expected for a convergent evolution. However, near
the blow-up time, the solutions diverge with higher resolutions providing
a larger derivative. The convergence factor $Q(t)$ is shown in the
middle frame; it likewise shows second-order convergence up to times
close to the blow-up time. In the bottom frame, the change in energy
$\Delta(t)$ is shown. We see that as the resolution is increased, energy
conservation improves.

What does this tell us about singularity formulation in wave maps evolved
from (supercritical) ingoing Gaussian pulse data?  We first argue that
these results are consistent with what would be expected for such
formation. As the singularity forms, higher and higher frequency
components become important, and they are represented numerically
only if one uses higher and higher grid  resolutions. Hence, the
behavior of the derivative of $\chi$ as the resolution improves would be
expected to show larger and larger gradients, as seen in Figure
2a. Next, we note that the formation of a singularity should not hinder
convergence except quite near the formation time, as is seen in Figure
2b.. Finally, energy conservation should be fine until the high frequency
components play their role, as we see in Figure 2c. Hence, the results
observed appear to be consistent with a singularity forming near $t=8$.

This does not guarantee that a singularity forms in these wave maps. There
are other effects that might produce the apparently unbounded growth of
the derivative of $\chi$ and of the energy density near the origin
in these numerical simulations. For example, perhaps some unphysical,
unstable mode  grows because of the particulars of our chosen evolution
scheme.  We believe that this is not the case, for a number of reasons.
First, the presence of such a mode would likely cause much larger
growth in $\Delta(t)$ than we see. Second, such modes would have to
be excited only after some time (roughly independent of resolution) and
only for families of sufficiently large energy. This is not consistent
with our observations. Third, the excitation of this sort of
instability would almost certainly depend critically on the precise
finite difference scheme. Because Bizo\'n and his
collaborators~\cite{bizon2+1} observe
similar behavior, using a different numerical evolution
scheme, this does not appear to be the case. Thus we believe it very
unlikely  that the effects we are seeing are the result of a numerically
unstable nonsingular mode.

Another situation in which one might numerically observe the formation of
singularities that do not in fact evolve analytically from the
corresponding data is if the continuum PDE solution is regularized by
high frequency components which cannot be seen by the finite grid
resolutions we use. The rather strong convergence behavior we see in our
numerical solutions leads us to believe that this is not happening. We
note in particular  that such unresolved components would have to be
separated in frequency space from the nontrivial low frequency components
by a substantial margin, with a large range of dynamically
irrelevant frequencies separating the two regimes. This seems to
be very unlikely.

\begin{figure}
\epsfxsize=8cm
\centerline{\epsffile{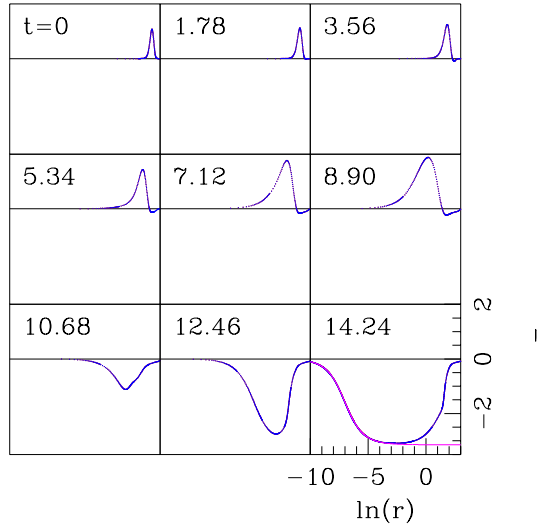}}
\caption{Near critical evolutions approach the static solution,
Eq.~(\ref{eq:static}).
         Nine frames equally spaced in time are shown for both
         {\em sub-} and {\em super-}critical evolutions. The solutions
         are indistinguishable in the graph at these
         times and are shown with dots. After $t=14.24$, the two
         solutions have quite different fates, but both approach
         the form of the static solution. In the final frame, the
         static solutions $\chi(r) = -2 \arctan \left( 1116 r \right)$
         is shown.
         }
\label{fig:crit_evol}
\end{figure}

\section{Conclusion}

The numerical studies we present here very strongly support the
contention, previously conjectured by Shatah and Struwe ~\cite{struwe},
that smooth initial data for wave maps from $2+1$ Minkowski spacetime
into the round two sphere can develop singularities (with unbounded
derivatives) in finite time. As we note, there are many ways in which the
numerical exploration of possible singularity formation might produce
misleading indications. However, we believe that as a consequence of the
numerical tests we have carried out, together with those done
independently by Bizo\'n and his collaborators ~\cite{bizon2+1}, the
formation of singularities is the most likely conclusion.

There is much more one would like to know about these
spatially equivariant wave maps, as well as about those without such
symmetry. One would like to know, for example, if the solutions assume any
universal form as one approaches the singularity. Our work
(see Fig.~\ref{fig:crit_evol})
supports the results of Bizo\'n et al~\cite{bizon2+1} which indicate that
indeed the
family of static spherically equivariant wave maps
\be
\chi(r) = \pm 2 \arctan \left( \lambda r \right)
\label{eq:static}
\ee
does serve as a sort of universal model for singularity formation. This
needs to be studied further. One would also very much like to understand
the behavior of the wave maps which evolve from initial data near the
transition from subcritical to supercritical data. The recent
numerical work of Bizo\'n et al ~\cite{bizon2+1} suggests that the static
solutions (\ref{eq:static}) play a central role in the evolution of the
transitional wave maps as well in that of supercritical ones;
however, this issue needs further investigation. (Note the
absence of any self-similar solutions to the 2+1 wave map equations; for
3+1 wave maps, such solutions play a key role in behavior of solutions
evolving from critical or near critical data).

\section*{Acknowledgments}

 We thank Piotr Bizo\'n  for helpful discussions.
Partial support for this work has come from NSF Grant
PHY-9800732 at the University of Oregon. SLL is appreciative of
the support of NSF PHY-9900644 and
of the financial support of Southampton College.


\end{document}